\documentclass[10pt,aps,prl,twocolumn,superscriptaddress]{revtex4}
\usepackage[linktocpage=true, colorlinks=true, urlcolor=blue, linkcolor=blue, citecolor=blue]{hyperref}
\usepackage{graphicx}
\usepackage{amsmath,amssymb}
\usepackage{color}
\usepackage{subfigure}
\usepackage{url}

\graphicspath{{../Images/}}

\begin{document}

\title{Experimental digital Gabor hologram rendering by a model-trained convolutional neural network}

\author{J. Rivet}
\author{A. Taliercio}
\author{C. Fang} 
\affiliation{
Centre National de la Recherche Scientifique (CNRS) UMR 7587, Institut Langevin. Paris Sciences et Lettres (PSL) Research University. Fondation Pierre-Gilles de Gennes, Institut National de la Sant\'e et de la Recherche M\'edicale (INSERM) U 979, Universit\'e Pierre et Marie Curie (UPMC), Universit\'e Paris 7. \'Ecole Sup\'erieure de Physique et de Chimie Industrielles ESPCI Paris - 1 rue Jussieu. 75005 Paris. France
}
\author{G. Tochon}
\author{T. G\'eraud}
\affiliation{
EPITA Research and Development Laboratory (LRDE), 14-16 rue Voltaire, F-94270 Le Kremlin-Bicetre, France
}
\author{JP. Huignard}
\author{M. Atlan}
\affiliation{
Centre National de la Recherche Scientifique (CNRS) UMR 7587, Institut Langevin. Paris Sciences et Lettres (PSL) Research University. Fondation Pierre-Gilles de Gennes, Institut National de la Sant\'e et de la Recherche M\'edicale (INSERM) U 979, Universit\'e Pierre et Marie Curie (UPMC), Universit\'e Paris 7. \'Ecole Sup\'erieure de Physique et de Chimie Industrielles ESPCI Paris - 1 rue Jussieu. 75005 Paris. France
}

\date{\today}

\begin{abstract} 
Digital hologram rendering can be performed by a convolutional neural network, trained with image pairs calculated by numerical wave propagation from sparse generating images. 512-by-512 pixel digital Gabor magnitude holograms are successfully estimated from experimental interferograms by a standard UNet trained with 50,000 synthetic image pairs over 70 epochs.
\end{abstract}

\maketitle


Convolutional neural networks already have demonstrated their potential for digital hologram rendering from optically-acquired interferograms in free-space propagation conditions~\cite{sinha2017lensless, horisaki2018single, rivenson2018phase, wang2018eholonet} and through scattering media~\cite{horisaki2016learning, li2018imaging, li2018deep}. Our aim here is to determine whether an auto-encoder convolutional neural network, a UNet~\cite{ronneberger2015u}, can be trained over a synthetic database for digital hologram rendering from experimental interferograms. A model of wave propagation is used to create synthetic Gabor interferograms and synthetic Gabor magnitude holograms from random images. This image formation model is based on angular spectrum propagation and magnitude calculation of the wave field from the object to the sensor array, and from the sensor to the object.\\

In contrast with previously reported computational image rendering schemes with convolutional neural networks, where image formation is statistically inferred through experimental data~\cite{rivenson2018phase, wang2018eholonet, horisaki2018single, sinha2017lensless}, in our approach it is inferred from synthetic data created by physical modeling of wave interference and propagation. Since the UNet training strategy relies on the strong use of a large and diverse database~\cite{ronneberger2015u}, training on synthetic data alleviates the need for numerous experimental data and data augmentation.\\


%
\begin{figure*}[]
\centering
\includegraphics[width =  \linewidth]{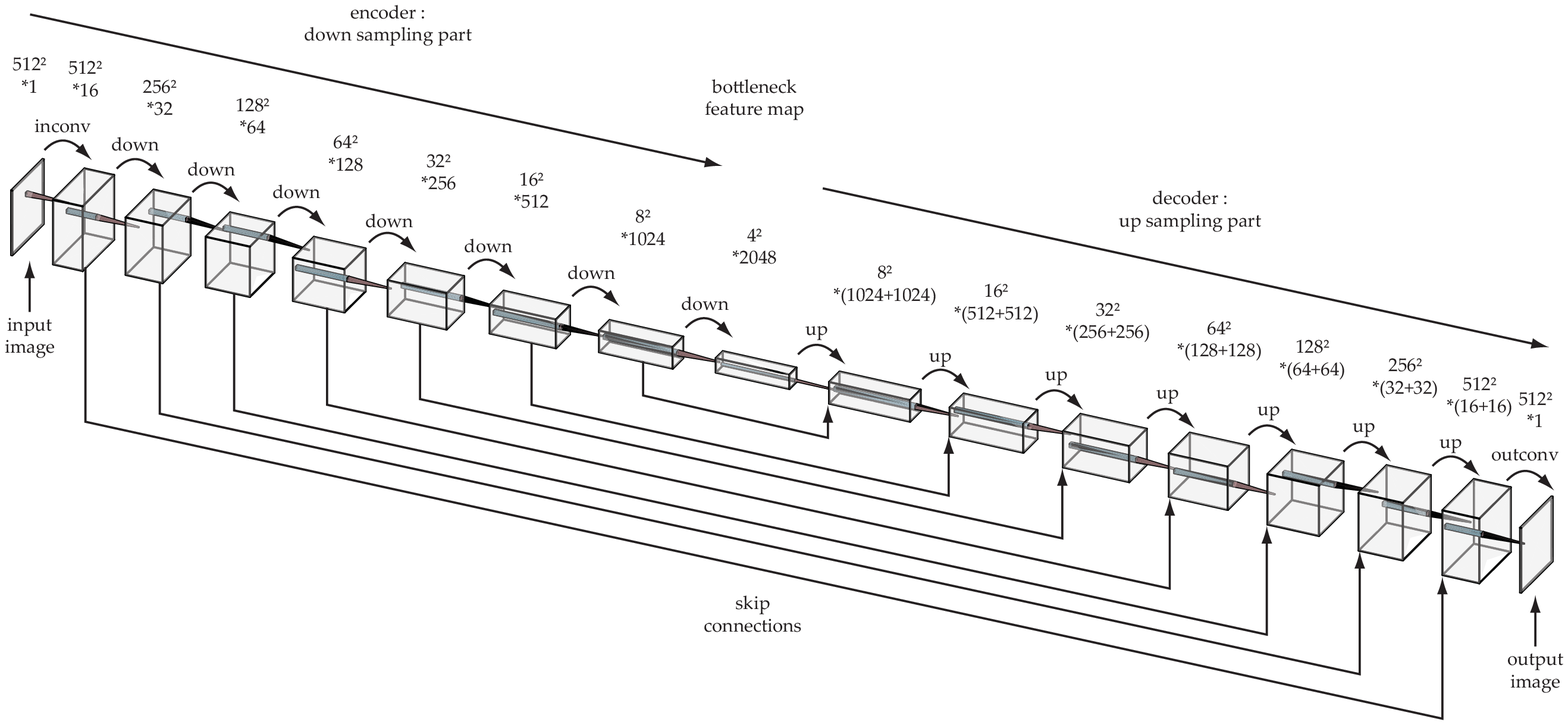}
\caption{Topology of the convolutional neural networks trained with synthetic, input/output image pairs, and used for experimental hologram rendering. Standard UNet~\cite{ronneberger2015u} with a positive real input (and output) image of $512^2$ pixels (width = height = 512 pixels), a depth of 7 down sampling blocks in the encoding part and 7 up sampling blocks in the decoding part. Convolution kernels are 3-by-3-by-$n$ pixels, where $n$ is the number of channels of the input feature map. The first set of kernels generates $n=16$ channels from the input image. In the down sampling part, the lateral size of the features is divided by two and the number of channels is multiplied by two between feature maps. Down sampling transitions (noted "down") include max pooling, and two iterations of convolution and rectification (ReLU)~\cite{glorot2011deep}. In the up sampling part, the lateral size of the features is multiplied by two and the number of channels is divided by two between blocks. Up sampling transitions (noted "up") include a convolution transpose, and two iterations of convolution and rectification. Mirror features from the down sampling part of the network are concatenated to their up sampling counterparts through skip connections that allow feature maps to pass through the bottleneck. The boxes represent feature maps, the numbers on top of each box are their respective width, height, and depth $n$. Flowchart courtesy of \href{http://alexlenail.me/NN-SVG/LeNet.html}{http://alexlenail.me/NN-SVG/LeNet.html}}
\label{fig_NeuralNetworkFlowChart}
\end{figure*}

The convolutional neural network used in this study is (sketched in Fig.~\ref{fig_NeuralNetworkFlowChart}) is a standard UNet~\cite{ronneberger2015u} with an input image of $512^2$ pixels, a depth of 7 down sampling blocks and 7 up sampling blocks. Convolution kernels are 3-by-3-by-$n$ pixels, where $n$ is the number of channels of the input feature map. The first set of 16 kernels generates a feature map of $n=16$ channels from the input image which has only $n=1$ channel. In the down sampling part, the lateral size of the features is divided by two and the number of channels $n$ is multiplied by two between blocks. In the up sampling part, the lateral size of the features is multiplied by two and the number of channels $n$ is divided by two between blocks. Mirror features from the down sampling part are concatenated to their up sampling counterparts. The UNet is trained with 50,000 image pairs (among which 15\% are used for validation purposes). The chosen loss function is the mean-square error between predicted image $H'$ and actual training output $H$ during the validation process. It is used to measure their inconsistency; the optimization (or deep learning) of the network consists in finding the set of network weights for which this loss function is minimum. The learning rate controls how much the weights of the network are adjusted with respect to the gradient of the loss function.\\


%
\begin{figure}[]
\centering
\includegraphics[width =  \linewidth]{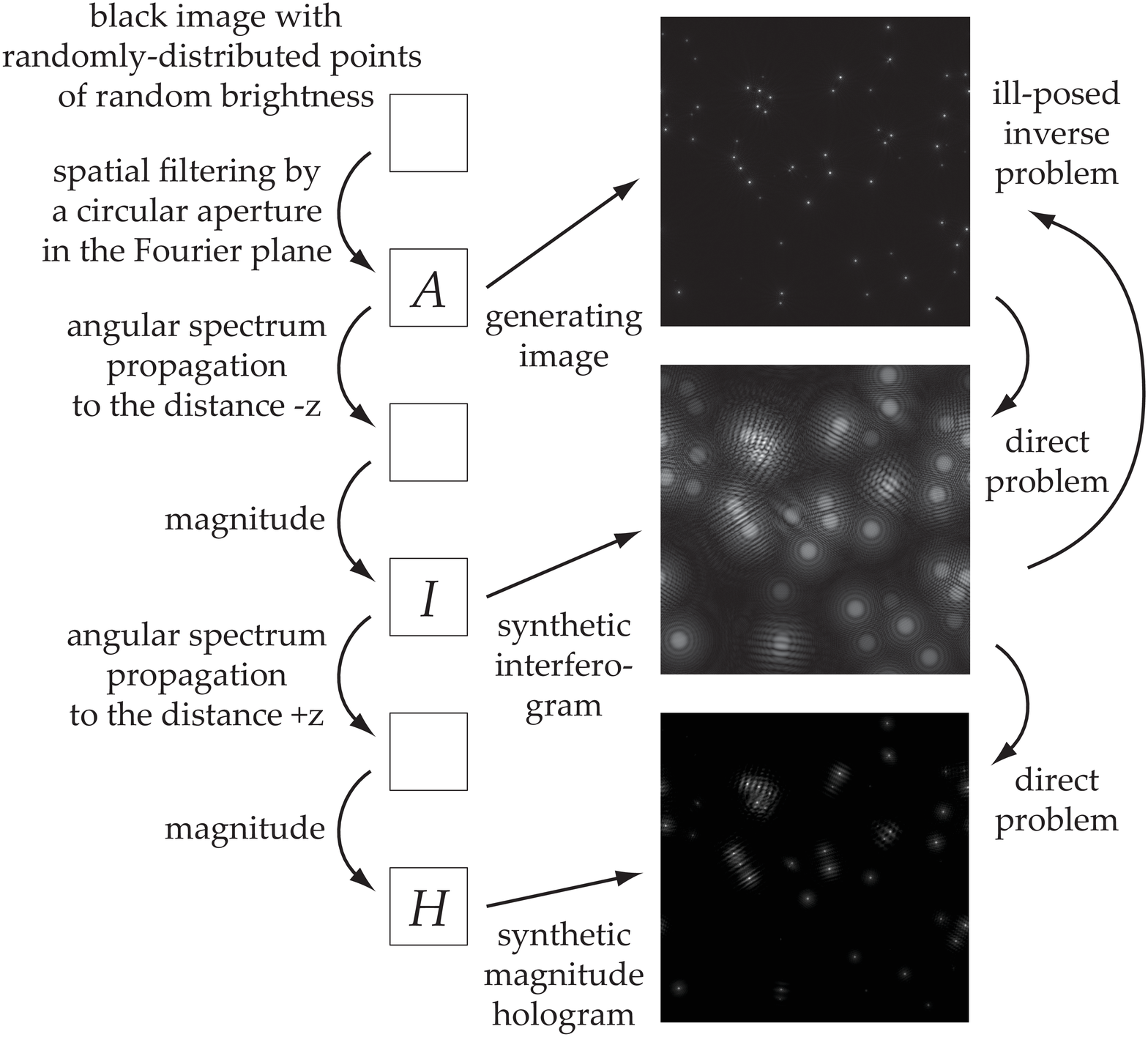}
\caption{Flowchart of the synthetic database creation. An image pair $(I,H)$ is calculated numerically from a random generating image $A$. This process is iterated for each random generating image to create the whole training database.}
\label{fig_DirectAndInverseProblemFlowchart}
\end{figure}

\begin{figure}[]
\centering
\includegraphics[width =  \linewidth]{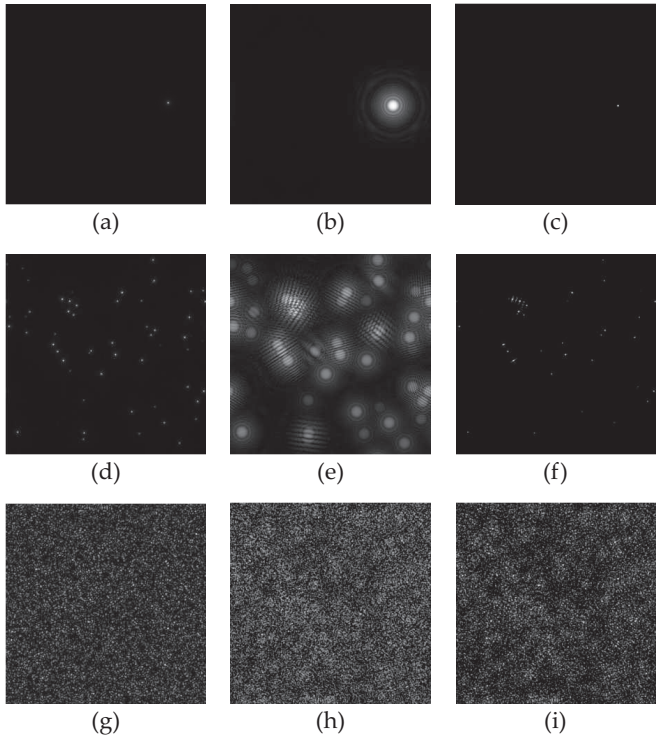}
\caption{Examples of database of image triplets $(A, I, H)$. Generating images $A$ (left column, a,d,g), synthetic interferograms $I$ (center column, b, e, h), synthetic magnitude holograms $H$ (right column, c, f, i). The number of random points in the generating images is $N=1$ (top row, a, b, c), $N=58$ (center row, d, e, f), $N=512^2/10$ (bottom row, g, h, i).  A movie of generated image triplets $(A,I,H)$ illustrating the distribution of the whole range of number of source points is displayed in \href{https://youtu.be/3pJCNV56ACI}{Vizualization 1}.}
\label{fig_NeuralNetworkTraining}
\end{figure}

We construct a database of training input and output image pairs by the procedure illustrated in the flowchart from Fig.~\ref{fig_DirectAndInverseProblemFlowchart}. A square generating image $A$ of $512^2$ pixels that describes the amplitude transmission function of a synthetic object is constructed by setting a given number $N$ of source points at random locations with random brightness on a black background, and spatial filtering by a circular aperture in the Fourier plane. The diameter of the aperture is one half of the diagonal of the reciprocal plane. The values of the array $A$ are positive real numbers. A synthetic Gabor interferogram $I$ is calculated from this generating image $A$ by angular spectrum propagation~\cite{goodman2008introduction} of the wave field described by the transmission function $A$ with a distance parameter $-z$, followed by a rectification consisting of taking the magnitude of the complex-valued array points
\begin{eqnarray}\label{eq_AngularSpectrumDiffractionIntegral1}
I(x,y) = \left| \iint {{\cal F}A}(k_x,k_y) e^{-i k_z z} e^{ik_x x} e^{ik_y y} {\rm d} k_x {\rm d} k_y \right|  
\end{eqnarray}
where $(x,y)$ are the pixels of arrays, and ${{\cal F}A}(k_x,k_y)$ is the two-dimensional Fourier transform of $A(x,y)$. The wave vector projections ($k_x$, $k_y$, and $k_z$) along lateral and axial directions ($x$, $y$, and $z$) satisfy $k_z^2 = k^2 - k_x^2 - k_y^2$, with $k = 2\pi / \lambda$, and $\lambda$ is the optical wavelength. A synthetic magnitude hologram $H$ is calculated from each synthetic interferogram $I$ by angular spectrum propagation of the wave field described by $I$ with a distance parameter $+z$, followed by rectification.
\begin{eqnarray}\label{eq_AngularSpectrumDiffractionIntegral2}
H(x,y) = \left| \iint {{\cal F}I}(k_x,k_y) e^{+i k_z z} e^{ik_x x} e^{ik_y y} {\rm d} k_x {\rm d} k_y \right|  
\end{eqnarray}
where ${{\cal F}I}(k_x,k_y)$ is the two-dimensional Fourier transform of $I(x,y)$. These operations generate a positive, real-valued image triplet $(A, I, H)$, displayed in Fig.~\ref{fig_NeuralNetworkTraining}. We ought to teach wave field propagation to a UNet, by deep learning over a large training database of $M$ randomly generated input/output image pairs $(I, H)$. The number of source points $N$ in each generating image $A$ is logarithmically-spaced from 1 to one-tenth of $512^2$.\\

%
 

By following the same construction procedure as for the generation of the training database, image couples $(I, H)$ are generated from a set of arbitrary images $A$ for validation purposes. The training procedure is stopped after 70 iterations of the optimization process over the whole training database (epochs), with a learning rate of 0.1, when the network output $H’$ for an input image $I$ becomes similar to the model-rendered magnitude hologram $H$.\\


%
\begin{figure}[]
\centering
\includegraphics[width =  \linewidth]{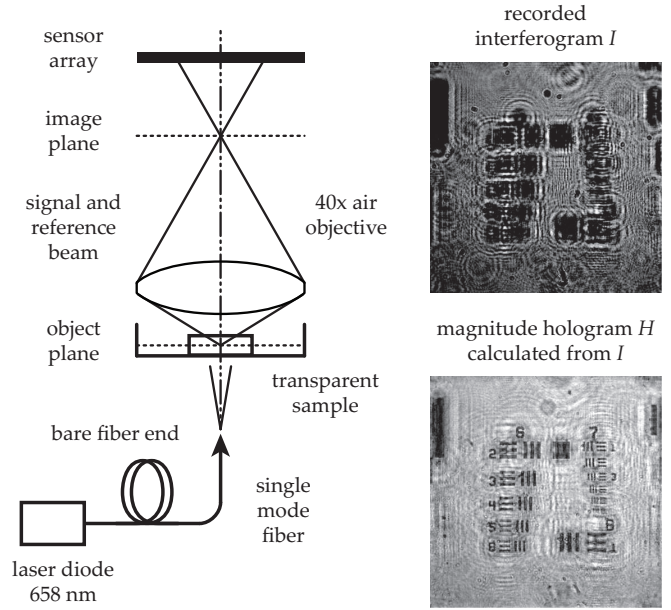}
\caption{Optical arrangement of the Gabor inline holographic microscope used to image transparent worms. A petri dish with growth medium placed in the object plane is illuminated by red laser light. It transmits light collected by a microscope objective, which creates an image conjugate of the sample in a plane between the lens and the sensor array of a camera. The recorded interferogram $I$ (top, right) of a resolution target placed in the object plane, yields the magnitude hologram $H$ (bottom, right) via Eq.~\ref{eq_AngularSpectrumDiffractionIntegral2}.}
\label{fig_SetupGaborMicroscope}
\end{figure}

\begin{figure}[]
\centering
\includegraphics[width = \linewidth]{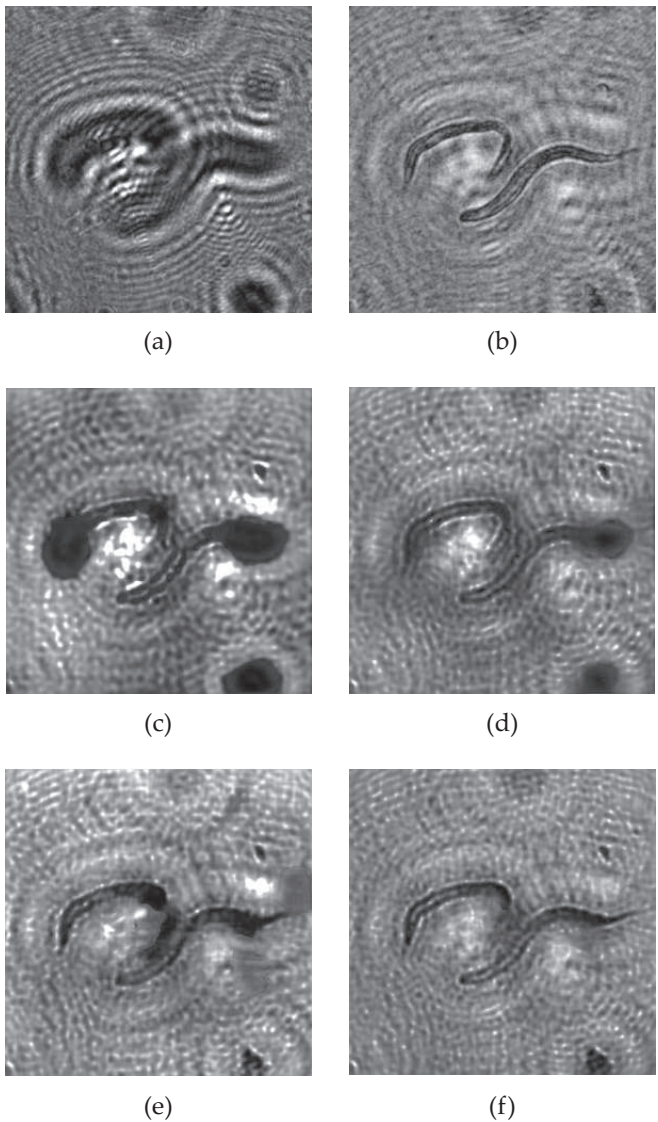}
\caption{(a) Optically-acquired digital interferogram $I$ of transparent worms. (b) Magnitude hologram $H$ calculated by angular spectrum propagation from $I$. Output image estimate $H'$ after 10 (c), 40 (d), 50 (e), and 70 (e) training epochs, over 50,000 synthetic image pairs $(I, H)$. A movie of image triplets $(I,H,H')$ is displayed in \href{https://youtu.be/77j8RZ-85EM}{Vizualization 2}.}
\label{fig_NeuralNetworkResults}
\end{figure}

Gabor interferograms $I$ measured from a preparation of C. elegans roundworms with a digital holographic Gabor microscope, sketched in Fig.~\ref{fig_SetupGaborMicroscope}, are then used to compare the network output $H'$ to magnitude holograms $H$ reconstructed by angular spectrum propagation (Eq.~\ref{eq_AngularSpectrumDiffractionIntegral2}). In the experiments, the radiation wavelength $\lambda$ is 658 nm, the pixel pitch is 5.5 microns, and the reconstruction distance is $z = 0.065$ m. 512-by-512-pixel interferograms $I$ are cropped from 2048-by-2048-pixel frames in a region of interest of the sensor array. A database of image couples $(I, H)$ is then constructed from a set of recorded Gabor interferograms $I$ and their magnitude hologram counterparts $H$, reconstructed by angular spectrum propagation from $I$, followed by rectification (Eq.~\ref{eq_AngularSpectrumDiffractionIntegral2}). Examples of network estimates $H’$ at several training iterations (epochs) for an input interferogram $I$, alongside the calculated magnitude hologram $H$ (Eq.~\ref{eq_AngularSpectrumDiffractionIntegral2}) are displayed in Fig.~\ref{fig_NeuralNetworkResults}. All the training dataset $(I,H)$ is calculated for $z = 0.065$ m. It is worth remarking that training the network over several reconstruction distances degrades the prediction accuracy.\\


%
%


Training the network with synthetic interferograms and reconstructed holograms $(I,H)$ image pairs teaches the network to estimate the solution of the transformation of Eq.~\ref{eq_AngularSpectrumDiffractionIntegral2}, for a given depth $z$, which already has an analytical solution. Yet this solution is cluttered with a spurious contribution. The quality of single-shot magnitude holograms reconstructed from Gabor in-line interferograms is degraded by the superposition of a twin image~\cite{denis2009inline, latychevskaia2007solution, zhang2018twin} : the ripples observed in the neighborhood of the worms in Fig.~\ref{fig_NeuralNetworkResults}(b) are the twin image of the roundworms in focus. The sum of the diffracted object wave beating against the reference wave, and their conjugate are present in the recorded interferogram $I$, hence the object wave reconstructed $+z$ is stained with an additive diffraction pattern, which creates a twin image at the reconstruction distance $-z$. Those ripples are also present in the image $H'$ in Fig.~\ref{fig_NeuralNetworkResults}(f), estimated by the neural network.\\

The convolutional neural network proves capable of mimicking standard hologram rendering with a high level of accuracy (Fig.~\ref{fig_NeuralNetworkResults}(f) vs. Fig.~\ref{fig_NeuralNetworkResults}(b)). We also wanted to assess wether it would also provide high quality estimates of solutions to the twin-image problem. Twin-image elimination by neural network rendering was previously reported for hologram estimation by a convolutional neural network~\cite{rivenson2018phase}. In this approach, the network was trained with interferograms measured experimentally and with calculated holograms from which the twin-image was removed by an experimental and numerical iterative multi-height phase recovery scheme~\cite{greenbaum2012maskless}. This suggests that UNets may be able to estimate solutions to ill-posed inverse problems beyond the ones for which the normal operator is a convolution~\cite{jin2017deep}. The inverse problem that needs to be solved is to determine the possible positive real-valued images (object amplitude transmission functions) to reproduce a given measured Gabor amplitude interferogram. Our network was also trained with $(I,A)$ pairs instead of $(I,H)$, ie. onto the inverse problem of image formation (Fig.~\ref{fig_DirectAndInverseProblemFlowchart}), switching the calculated magnitude holograms $H$ for generating images $A$, naturally devoid of twin image. Yet it did not enable the neural network to estimate twin-image-free magnitude holograms $H'$ from inline interferograms inputs $I$. This approach failed to reconstruct twin-image-free Gabor holograms. This problem is most often ill-posed, which means that many object transmission estimates may produce the same Gabor amplitude interferogram. Yet the direct problem, which is the formation of an interferogram by a given transmission function, has an analytical formulation. Adding regularization constraints \cite{fienup1978reconstruction} has emerged as the standard procedure for iterative image reconstruction algorithms~\cite{denis2009inline, latychevskaia2007solution, jin2017deep, zhang2018twin}. It may be also prove useful for hologram rendering by convolutional networks.\\

In conclusion, digital image rendering in Gabor holography can be performed by a convolutional neural network trained with a fully synthetic database formed by image pairs generated randomly, and linked by a numerical model of in-line angular spectrum propagation of a scalar wave field from the object to the sensor array, and magnitude calculation. Gabor holograms of microscopic worms are successfully predicted from experimental interferograms by a UNet trained with 50,000 random image pairs. Two main caveats apply to the use of a standard Unet for image rendering : the results were obtained for a fixed reconstruction distance, and twin-image elimination could not be achieved by training the network with image pairs from the inverse problem.\\


This work was supported by LABEX WIFI (Laboratory of Excellence ANR-10-LABX-24) within the French Program Investments for the Future under Reference ANR-10-IDEX-0001-02 PSL, and European Research Council (ERC Synergy HELMHOLTZ, grant agreement \#610110). The Titan Xp used for this research was donated by the NVIDIA Corporation.\\


We are thankful to Vincent Galy for providing the round worms C. elegans, and we acknowledge valuable assistance from Armelle Rancillac, Nicolas Letort and St\'ephanie Rind.


\end{document}